\documentclass[11pt,fleqn]{article}

\usepackage{amsmath,amssymb,amsthm,enumerate,lscape}%,cite,backref

\newtheorem*{proposition*}{Proposition}
{\theoremstyle{definition}

}

\newcommand{\todo}[1][\null]{\ensuremath{\clubsuit}}

\newcommand{\noprint}[1]{}

\begin{document}

\par\noindent {\LARGE\bf
Extended symmetry analysis\\ of a ``nonconservative Fokker--Plank equation''
\par}

{\vspace{2mm}\par\noindent
Vyacheslav BOYKO~$^{\dag}$ and Nataliya SHAPOVAL~$^{\ddag}$
\par\vspace{1mm}\par}

{\vspace{2mm}\par\noindent\it
${}^\dag$~Institute of Mathematics of NAS of Ukraine, 3 Tereshchenkivska Str., Kyiv-4, Ukraine\\
\hphantom{${}^\dag$}~\rm E-mail: \it  boyko@imath.kiev.ua
\\[1mm]
${}^\ddag$~Faculty of Mechanics and Mathematics, National Taras Shevchenko University of Kyiv,\\
\hphantom{${}^\ddag$}~2 Academician Glushkov Ave., 03127 Kyiv, Ukraine
\\
\hphantom{${}^\ddag$}~\rm E-mail: \it  natalya.shapoval@gmail.com
\par}

{\vspace{3mm}\par\noindent\hspace*{8mm}\parbox{140mm}{\small
We show that all results of Ya\c{s}ar and \"{O}zer [{\it Comput. Math. Appl.} {\bf 59} (2010), 3203--3210]
on symmetries and conservation laws of a ``nonconservative Fokker--Planck equation''
can be easily derived from results existing in the literature
by means of using the fact that this equation is reduced to the linear heat equation by a simple point transformation.
Moreover nonclassical symmetries and local and potential conservation laws of the equation under consideration
are exhaustively described in the same way as well as infinite series of potential symmetry algebras of arbitrary potential orders
are constructed.\\[1ex]
}\par\vspace{4mm}}

The investigation of Lie symmetries of two-dimensional second-order linear partial differential equations
was one of the first problems considered within the group analysis of differential equations.
The complete group classification of such equations was carried out by Lie~\cite{Lie1881} himself
and essentially later revisited by Ovsiannikov~\cite{Ovsiannikov1982}.
This classification includes, as a special case, the group classification of
linear $(1+1)$-dimensional homogeneous second-order evolution equations of the general form
\begin{gather}\label{EqGenLPE}
u_t=a(t,x)u_{xx}+b(t,x)u_x+c(t,x)u,
\end{gather}
where $a$, $b$ and $c$ are arbitrary smooth functions of~$t$ and~$x$, $a\ne0$.
On the other hand the study of point equivalences between linear evolution equations also have a long history.
It was started with Kolmogorov's paper~\cite{Kolmogorov1938}, where the problem of the description of Kolmogorov equations that
are reduced to the linear heat equation by certain point transformations was posed.
This problem was exhaustively solved by Cherkasov~\cite{Cherkasov1957}.
Symmetry criteria on equivalence of equations within class~\eqref{EqGenLPE} obviously follow from the above Lie--Ovsiannikov group classification. They were also discussed by a number of authors \cite{Bluman1980,Shtelen&Stogny1989,Spichak&Stognii1999c}.
See additionally the review in~\cite{PopovychKunzingerIvanova2008} and references therein.

A constructive approach to investigation of point equivalences of equations from the class~\eqref{EqGenLPE} was suggested by Ibragimov \cite{Ibragimov2002}.
This approach is based upon the computation of differential invariants and semi-invariants of the associated equivalence group~$G^\sim$.
In fact in \cite{Ibragimov2002} only the semi-invariants of the subgroup of~$G^\sim$,
consisting of linear transformations in the dependent variable, were obtained.
An efficient criterion on point equivalence of equations from class~\eqref{EqGenLPE} to the linear heat equation in terms of
semi-invariants of the entire equivalence group~$G^\sim$, depending upon the coefficients of equation~\eqref{EqGenLPE}, was proposed in~\cite{Johnpillai&Mahomed2001}.
This criterion was reformulated in a more compact form in \cite{Mahomed2008}.
Therein criteria of point reducibility to Lie's other canonical forms were found within Ibragimov's approach.
It is necessary to note that earlier the equivalence problem for equations from class~\eqref{EqGenLPE}
was exhaustively investigated by Morozov~\cite{Morozov2003} within the framework of the method of moving frames
although the expressions constructed for related differential invariants and semi-invariants are quite cumbersome.

An exhaustive review of the previous results on group analysis and the equivalence problem within the class~\eqref{EqGenLPE} as well as the complete description of conservation laws and potential symmetries of equations from this class can be found in the paper by Popovych, Kunzinger and Ivanova~\cite{PopovychKunzingerIvanova2008}.

In a recent paper \cite{Yasar2010} %Ya\c{s}ar and \"{O}zer
Lie symmetries, conservation laws, potential symmetries and solutions were considered for the ``nonconservative Fokker--Planck equation''
\begin{gather}\label{my_eq1}
u_t=u_{xx}+xu_x
\end{gather}
which is a representative of the class~\eqref{EqGenLPE}.
After the sign of~$t$ is changed, this equation becomes a Kolmogorov backward equation.
Although equation~\eqref{my_eq1} is quite interesting from the mathematical point of view and appears in numerous applications,
its investigation in the framework of symmetry analysis is unnecessary due to the fact that it is reduced to the linear heat equation
\begin{gather}\label{my_eq3}
w_\tau=w_{yy}
\end{gather}
by the simple point transformation
\begin{gather}\label{my_eq2}
\tau=\frac{1}{2}e^{2t}, \quad y=e^tx, \quad w=u.
\end{gather}
The inverse of \eqref{my_eq2},
\begin{gather}\label{my_eq2'}
t=\frac12\ln(2\tau),\quad x=\frac{y}{\sqrt{2\tau}}, \quad u=w,
\end{gather}
is well defined  only for $\tau>0$.

Because of the transformation \eqref{my_eq2} the results of \cite{Yasar2010} as well as more general ones can be easy derived from well-known results on the linear heat equation~\eqref{my_eq3} existing in the literature for a long time, namely, using the transformation \eqref{my_eq2}, one can obtain the maximal Lie invariance algebra of the  equation~\eqref{my_eq1}, %(cf.\ Section~3 of \cite{Yasar2010}),
the complete description of its local and potential conservation laws %(cf.\ Section~4 of \cite{Yasar2010})
and its potential symmetries of any potential order  %(cf.\ Section~5 of \cite{Yasar2010})
as well as nonclassical symmetries and wide families of its exact solutions.
(A similar example on usage of point equivalence of a Fokker--Planck equation to the linear heat equation was presented in~\cite[p.~156]{PopovychKunzingerIvanova2008}.)

The maximal Lie invariance algebra of the linear heat equation~\eqref{my_eq3} is
\begin{gather*}%\label{my_eq4}
\mathfrak g=\langle
\partial_\tau,\, \partial_y,\, 2\tau\partial_\tau+y\partial_y,\, 2\tau\partial_y-yw\partial_w,\, 4\tau^2\partial_\tau+4\tau y\partial_y-(y^2+2\tau)w\partial_w,\, w\partial_w,\, f\partial_w\rangle,
\end{gather*}
where the function $f=f(\tau,y)$ is an arbitrary solution of~\eqref{my_eq3}~\cite{Lie1881,Olver1993,Ovsiannikov1982}.
The maximal Lie invariance algebra $\tilde{\mathfrak g}$ of the equation~\eqref{my_eq1} was calculated in Section~3 of~\cite{Yasar2010}  by the classical infinitesimal method.
In contrast to this we directly obtain~$\tilde{\mathfrak g}$ from~$\mathfrak g$ using the pushforward of vector fields associated with~\eqref{my_eq2'}:
\begin{gather*}
\tilde{\mathfrak g}=\langle\partial_t,\,  e^{-t}\partial_x,\, e^{-2t}\partial_t-e^{-2t}x\partial_x+e^{-2t}u\partial_u,\,  e^t\partial_x-e^txu\partial_u,
 \\
 \hphantom{\tilde{\mathfrak g}=\langle}{} e^{2t}\partial_t+e^{2t}x\partial_x-e^{2t}x^2u\partial_u,\,  u\partial_u,\,  f\partial_u \rangle,
\end{gather*}
where the function $f=f(t,x)$ runs through the solution set of~\eqref{my_eq1}.

In emulation of Ibragimov's approach \cite{Ibragimov2007},
in Section~4 of~\cite{Yasar2010} conservation laws of the joint system of the initial equation~\eqref{my_eq1} and its adjoint
\begin{gather}\label{my_eq1-ad}
\alpha_t+\alpha_{xx}-(x\alpha)_x=0
\end{gather}
were constructed using Lie symmetries of the system,
namely essential Lie symmetries of~\eqref{my_eq1} were prolonged to the additional dependent variable~$\alpha$.
The prolonged symmetries are Lie symmetries of the joint system of~\eqref{my_eq1} and~\eqref{my_eq1-ad}
and variational symmetries of the corresponding Lagrangian $\alpha(u_t-u_{xx}-xu_x)$.
Therefore conserved vectors of the system are obtained from the prolonged symmetries
in view of the Noether theorem on connections between symmetries and conservation~laws.

All conservation laws of the joint system of the linear heat equation~\eqref{my_eq3} and its adjoint (the linear backward heat equation),
\begin{gather}\label{my_eq3-ad}
\tilde \alpha_\tau=-\tilde \alpha_{yy},
\end{gather}
obtainable with this approach, were calculated in \cite{Ibragimov2007,IbragimovKolsrud2004}.
The point transformation~\eqref{my_eq2}  prolonged to the dependent variable $\alpha$ according to the formula $\tilde \alpha=e^{-t}\alpha$
(cf.\ Section~5 of~\cite{PopovychKunzingerIvanova2008} and in particular equation~(26))
maps the joint system of~\eqref{my_eq1} and~\eqref{my_eq1-ad} to the joint system of~\eqref{my_eq3} and~\eqref{my_eq3-ad}.
Therefore the result of Section~4 of~\cite{Yasar2010}
is a direct consequence of the result for linear heat equation from~\cite{Ibragimov2007,IbragimovKolsrud2004}.

The complete description of local and potential conservation laws of~\eqref{my_eq1} can be simply obtained
as a particular case of the same results of~\cite{PopovychKunzingerIvanova2008}
on the entire class of linear $(1+1)$-dimensional second-order evolution equations~\eqref{EqGenLPE}.
According to \cite[Lemma~3]{PopovychKunzingerIvanova2008}
every local conservation law of any equation from class~\eqref{EqGenLPE} is of the first order
and moreover it possesses a conserved vector with density depending at most upon $t$, $x$ and $u$
and flux depending at most on $t$, $x$, $u$ and~$u_x$.
As was proven in~\cite[Theorem~4]{PopovychKunzingerIvanova2008}, % p.~133
each local conservation law of an arbitrary equation from the class~\eqref{EqGenLPE} contains
a conserved vector of the canonical form
\begin{gather*}%\label{eqCVofLPEs}
\bigl(\alpha u,\, -\alpha au_x+((\alpha a)_x-\alpha b)u\bigr),
\end{gather*}
where the characteristic $\alpha=\alpha(t,x)$ runs through the solution set of the adjoint equation
\begin{gather}\label{EqAdjLPE}
\alpha_t+(a\alpha)_{xx}-(b\alpha)_x+c\alpha=0.
\end{gather}
In other words the space of local conservation laws of equation~\eqref{EqGenLPE} is isomorphic
to the solution space of the adjoint equation~\eqref{EqAdjLPE}.
An analogous statement is true for an arbitrary $(1+1)$-dimensional linear evolution equation of even order~\cite{PopovychSergyeyev2010}.

As a result the equation~\eqref{my_eq1} possesses, up to the equivalence of conserved vectors~\cite[Definition~3]{PopovychKunzingerIvanova2008},
only conserved vectors of the canonical form
\[
\bigl(\alpha u,\, -\alpha u_x+(\alpha_x-\alpha x)u\bigr),
\]
where the characteristic $\alpha=\alpha(t,x)$ is an arbitrary solution of the adjoint equation~\eqref{my_eq1-ad}.
Moreover it follows from Theorem~5 of~\cite{PopovychKunzingerIvanova2008} that any potential conservation law of equation~\eqref{my_eq1}
is induced by a local conservation law of this equation.

Potential symmetries of~\eqref{my_eq1} also can be constructed from known potential symmetries of the linear heat equation~\eqref{my_eq3}.

It was proven in \cite[Theorem~7]{PopovychKunzingerIvanova2008} that
the linear heat equation~\eqref{my_eq3} admits, up to the equivalence generated by its point symmetry group, only two simplest potential systems
the Lie symmetries of which are not induced by Lie symmetries of~\eqref{my_eq3} and therefore are nontrivial potential symmetries of~\eqref{my_eq3}.
(A potential system is called simplest if it involves a single potential.)
These potential systems are associated with the characteristics~$\alpha^1=1$ and $\alpha^2=y$.
The inverse
\begin{gather*}%\label{my_eq2-prol}
t=\frac12\ln(2\tau),\quad x=\frac{y}{\sqrt{2\tau}}, \quad u=w, \quad \alpha=e^t\tilde\alpha
\end{gather*}
of the transformation~\eqref{my_eq2} prolonged to the dependent variable $\alpha$ maps
the characteristics~$\alpha^1$ and $\alpha^2$ of equation~\eqref{my_eq3} to
the characteristics $\tilde{\alpha}^1=e^t$ and $\tilde{\alpha}^2=e^{2t}x$ of equation~\eqref{my_eq1}.
The potential systems for equation~\eqref{my_eq1}, associated with the characteristics~$\tilde{\alpha}^1$ and~$\tilde{\alpha}^2$, respectively are
\begin{gather}\label{pot1FP}
\hat v_x= e^tu,\quad
\hat v_t= e^tu_ x+ e^t xu\\
\intertext{and}
\label{pot2FP}
\check v_x=e^{2t}xu, \quad
\check v_t=e^{2t}xu_x+e^{2t}(x^2-1)u.
\end{gather}

Their maximal Lie invariance algebras are
\begin{gather*}
\tilde{\mathfrak p}_1=\langle\ e^{-2t}\partial_t-e^{-2t}x\partial_x,
\ e^{-t}\partial_x,\
\partial_t-u\partial_u,\
e^t\partial_x-e^t(xu+x\hat v)\partial_u-\hat v\partial_{\hat v},
\\ \phantom{\tilde{\mathfrak p}_1=\langle\ }
e^{2t}\partial_t+e^{2t}x\partial_x-e^{2t}(x^2u+3u+2xe^{-t} \hat v)\partial_u-e^{2t}(x^2+1)\hat v\partial_{\hat v},
\\ \phantom{\tilde{\mathfrak p}_1=\langle\ }
 u\partial_u+\hat v\partial_{\hat v}, \
e^{-t}g_x\partial_u+g\partial_{\hat v}\, \rangle,
\\[1ex]
\tilde{\mathfrak p}_2=\langle\
\partial_t-u\partial_u,\
e^{-2t}\partial_t-e^{-2t}x\partial_x,\
\\ \phantom{\tilde{\mathfrak p}_2=\langle\ }
e^{2t}\partial_t+e^{2t}x\partial_x-e^{2t}(x^2u+3u+2e^{-2t}\check v)\partial_u-e^{2t}(x^2-1)\check v\partial_{\check v},
\\ \phantom{\tilde{\mathfrak p}_2=\langle\ }
u\partial_u+\check v\partial_{\check v},\
e^tx^{-1}h_x\partial_u+h\partial_{\check v}\, \rangle,
\end{gather*}
where the functions $g=g(t,x)$ and $h=h(t,x)$ run through the solution set of the associated potential equations,
$\hat v_t-\hat v_{xx}-x\hat v_x=0$
and
$\check v_t-\check v_{xx}+(2x^{-1}-x)\check v_x=0$, respectively.
The maximal Lie invariance algebras of these potential equations are projections
of~$\tilde{\mathfrak p}_1$ and $\tilde{\mathfrak p}_2$ to the spaces $(t,x,\hat v)$ and $(t,x,\check v)$.
Instead of using the infinitesimal Lie method,
the algebras $\tilde{\mathfrak p}_1$ and $\tilde{\mathfrak p}_2$ can be obtained from
the potential symmetry algebras $\mathfrak p_1$ and $\mathfrak p_2$ of the linear heat equation
(see~\cite[pp.~155--156]{PopovychKunzingerIvanova2008}) by the transformation~\eqref{my_eq2'}
trivially prolonged to the corresponding potentials.

Moreover the linear heat equation admits an infinite series $\{\mathfrak g_p,\, p\in\mathbb N\}$ of potential symmetry algebras
\cite[Proposition 12]{PopovychKunzingerIvanova2008},
namely, for any $p\in\mathbb N$ the algebra $\mathfrak g_p$ is of strictly $p$th potential order
(i.e., it involves exactly $p$ independent potentials)
and is associated with $p$-tuples of characteristics
which are linearly independent polynomial solutions of lowest order of the backward heat equation.
Each of the algebras $\mathfrak g_p$ is isomorphic to the maximal invariance algebra $\mathfrak g$ of the linear heat equation.
The linear heat equation possesses also other infinite series of nontrivial potential symmetry algebras.
Due to the change of variables~\eqref{my_eq2'}, trivially prolonged to the corresponding potentials,
similar results hold true for the equation~\eqref{my_eq1}.

In \cite{Yasar2010} the classical Lie algorithm was used for finding the maximal Lie invariance algebra
of the potential system of the equation~\eqref{my_eq1}, which is associated with the characteristic~$e^{-x^2/2}$.
In fact this system is equivalent, with respect to the Lie symmetry group of~\eqref{my_eq1},
to the simpler potential system~\eqref{pot1FP} associated with the characteristic~$\tilde{\alpha}^1=e^t$.
To reduce the characteristic~$e^{-x^2/2}$ to the characteristic~$\tilde{\alpha}^1$ it is necessary to apply
a point symmetry transformation which is the composition of a projective transformation and a shift with respect to~$t$.

Nonclassical symmetries of the linear heat equation~\eqref{my_eq3} were exhaustively investigated in~\cite{Fushchych&Shtelen&Serov&Popovych1992}.
Roughly speaking, it was proved that in both the singular and regular cases
(when the coefficient of~$\partial_t$ in a nonclassical symmetry operator vanishes or does not, respectively)
the corresponding determining equations are reduced by nonpoint transformations to the initial equations.
Later these no-go results were extended to all equations from the class~\eqref{EqGenLPE} \cite{Popovych2006b,Popovych2008,Zhdanov&Lahno1998}.

Continuing the list of common errors in finding exact solutions of differential equations made up by Kudryashov~\cite{Kudryashov2009},
Popovych and Vaneeva~\cite{PopovychVaneeva2010} indicated one more common error of such a kind:
Solutions are often constructed with no relation to equivalence of differential equations
with respect to point (resp.\ contact, resp.\ potential etc.) transformations.
A number of multiparametric families of exact solutions of the linear heat equation~\eqref{my_eq3} are well known for a long time
and are presented widely in the literature.
See e.g.\ the review~\cite{Ivanova2008}, the textbook~\cite[Examples 3.3 and 3.17]{Olver1993},
the handbook \cite{Polyanin2002} and the website EqWorld {\tt http://eqworld.ipmnet.ru/}.
A simple solution of~\eqref{my_eq1} was constructed in Section~5 of~\cite{Yasar2010}
which is similar with respect to the point transformation~\eqref{my_eq2}
to the obvious solution $w=c_1y$ of the linear heat equation~\eqref{my_eq3}.

\medskip

{\bf Acknowledgements.}
The authors are grateful to Roman Popovych for helpful discussion and remarks as well as for relevant references.
The research of V.\,B. was supported by the Austrian Science Fund (FWF), project P20632.
V.\,B. is grateful for the hospitality by the Department of Mathematics of the University of Vienna.


\begin{thebibliography}{200}
\small\itemsep=-1pt

\bibitem{Bluman1980}
Bluman G.W.,
On the transformation of diffusion processes into the Wiener process,
{\it SIAM J. Appl. Math.} {\bf  39} (1980), 238--247.

\bibitem{Cherkasov1957}
Cherkasov~I.D., On the transformation of the diffusion process to a Wiener process,
{\it Theory Probab. Appl.} {\bf 2} (1957), 373--377.

\bibitem{Fushchych&Shtelen&Serov&Popovych1992}
Fushchych W.I., Shtelen W.M., Serov M.I., Popovych R.O.,
$Q$-conditional symmetry of the linear heat equation,
{\it Proc. Acad. of Sci. Ukraine} (1992), no.~12, 28--33.

\bibitem{Ibragimov2002}
Ibragimov~N.H.,
Laplace type invariants for parabolic equations,
{\it Nonlinear Dynam.} {\bf 28} (2002), 125--133.

\bibitem{Ibragimov2007}
Ibragimov N.H.,
A new conservation theorem,
{\it J. Math. Anal. Appl.} {\bf 333} (2007), 311--328.

\bibitem{IbragimovKolsrud2004}
Ibragimov N.H., Kolsrud T.,
Lagrangian approach to evolution equations: symmetries and conservation laws,
{\it Nonlinear Dynam.} {\bf  36} (2004), 29--40.

\bibitem{Ivanova2008}
Ivanova N.M.,
Exact solutions of diffusion-convection equations,
{\it Dynam. PDEs} {\bf 5} (2008), 139--171.

\bibitem{Johnpillai&Mahomed2001}
Johnpillai I.K., Mahomed F.M.,
Singular invariant equation for the $(1+1)$ Fokker--Planck equation,
{\it J. Phys. A: Math. Gen.} {\bf 34} (2001), 11033--11051.

\bibitem{Kolmogorov1938}
Kolmogorov A.N.,
On analytic methods in probability,
{\it Uspehi Mat. Nauk} {\bf 5} (1938), 5--41.

\bibitem{Kudryashov2009}
Kudryashov N.A.,
Seven common errors in finding exact solutions of nonlinear differential equations,
{\it Commun. Nonlinear Sci. Numer. Simul.} {\bf 14} (2009), 3507--3529.

\bibitem{Lie1881}
Lie S.,
\"Uber die Integration durch bestimmte Integrale von einer Klasse linear partieller
Differentialgleichung, {\it Arch. Math.} {\bf 6} (1881), no.~3, 328--368
(translation by N.H. Ibragi\-mov:
Lie S., On integration of a class of linear partial differential equations by means of
definite integrals, {\it CRC Handbook of Lie Group Analysis of Differential Equations},
Vol. 2, 1994, 473--508).

\bibitem{Mahomed2008}
Mahomed F.,
Complete invariant characterization of scalar linear $(1+1)$ parabolic equations,
{\it J.~Nonlinear Math. Phys.} {\bf  15} (2008), 112--123.

\bibitem{Morozov2003}
Morozov O.I.,
Contact equivalence problem for linear parabolic equations,
math-ph/0304045.

\bibitem{Olver1993}
Olver P.J.,
{\it Application of Lie groups to differential equations},
Springer-Verlag, New York, 1993.

 \bibitem{Ovsiannikov1982}
Ovsiannikov~L.V.,
{\it Group analysis of differential equations},
Academic Press, New York, 1982.

\bibitem{Polyanin2002}
Polyanin A.D.,
{\it Handbook of linear partial differential equations for engineers and scientists},
Chapman \& Hall/CRC, Boca Raton, FL, 2002.

\bibitem{Popovych2006b}
Popovych R.O.,
No-go theorem on reduction operators of linear second-order parabolic equations,
in {\it Symmetry and Integrability of Equations of Mathematical Physics}, {\it Collection of Works of Institute of Mathematics, Kyiv} {\bf 3} (2006), no.~2, 231--238.

\bibitem{Popovych2008}
Popovych R.O.,
Reduction operators of linear second-order parabolic equations,
{\it J. Phys. A: Math. Theor.} {\bf 41} (2008), 185202, 31~pages.

\bibitem{PopovychKunzingerIvanova2008}
Popovych R.O., Kunzinger M., Ivanova N.M.,
Conservation laws and potential symmetries of linear parabolic equations,
{{\it Acta. Appl. Math.}} {\bf 100} (2008), 113--185.

\bibitem{PopovychSergyeyev2010}
Popovych R.O., Sergyeyev A.,
Conservation laws and normal forms of evolution equations,
{\it Phys. Lett. A} {\bf 374} (2010), 2210--2217.

\bibitem{PopovychVaneeva2010}
Popovych R.O., Vaneeva O.O.,
More common errors in finding exact solutions of nonlinear differential equations.~I,
{\it Commun. Nonlinear Sci. Numer. Simul.} {\bf 15} (2010), 3887--3899.

\bibitem{Shtelen&Stogny1989}
Shtelen W.M., Stogny V.I.,
Symmetry properties of one- and two-dimensional Fokker--Planck equations,
{\it J. Phys. A: Math. Gen.} {\bf 22} (1989), L539--L543.

\bibitem{Spichak&Stognii1999c}
Spichak S.V., Stognii V.I.,
Symmetry classification and exact solutions of the one-dimensional Fokker--Planck equation with arbitrary coefficients of drift and diffusion,
{\it J.~Phys.~A: Math. Gen.} {\bf 32} (1999), 8341--8353.

\bibitem{Yasar2010}
Ya\c{s}ar E., \"{O}zer T.,
Invariant solutions and conservation laws to nonconservative FP equation,
{\it Comput. Math. Appl.} {\bf 59} (2010), 3203--3210.

\bibitem{Zhdanov&Lahno1998}
Zhdanov R.Z., Lahno V.I.,
Conditional symmetry of a porous medium equation,
{\it Phys. D} {\bf 122} (1998), 178--186.

\end{thebibliography}
\end{document}